\documentstyle[preprint,aps]{revtex}
\begin{document}

\begin{flushright} IPNO/TH 93-38 \end{flushright}

\vspace{2cm}
\begin{center}
{\Large {\bf Light Quark Masses from Exclusive Tau Decays: \\
An Experimental Proposal}}

\indent

J. Stern, N.H. Fuchs\footnote{Permanent address:
Department of Physics, Purdue University, West Lafayette IN 47907 USA.}
and M. Knecht\\

\indent

{\sl Division de Physique Th\'{e}orique
\footnote{Unit\'{e} de Recherche des
Universit\'{e}s Paris XI et Paris VI associ\'{e}e au CNRS}\\
Institut de Physique Nucl\'{e}aire\\
F-91406 Orsay Cedex, France}

\end{center}

\vspace{4cm}

\centerline{\bf Abstract}

\indent

A method of indirect measurement of the light quark running mass
$\hat{m} = (m_d + m_u)/2$ is elaborated in detail.  It is based on measuring
1\%-level azimuthal angular asymmetries in the decay $\tau \rightarrow
\nu_\tau + 3\pi$.  The latter are then used in QCD sum rules to obtain
experimental lower bounds for $\hat{m}$.  For a sample of $2.5 \times 10^5$
$\tau \rightarrow \nu_\tau + 3\pi$ decays free of background, the resulting
statistical error in the bound for $\hat{m}$ is estimated to be 1 MeV, i.e.,
comparable to the systematic error due to the use of QCD sum rules.

\vspace{2cm}

{\em Contribution to the Third Workshop on the $\tau$ Charm Factory\\
1-6 June 1993, Marbella, Spain}

\pagebreak
{\flushleft{\bf 1. INTRODUCTION}}

The QCD Lagrangian contains seven parameters (apart from the vacuum
angle $\theta$) which are not determined by theoretical
considerations: The gauge coupling constant $g$ and the quark masses
$m_a$, one for each flavor ($a=u,d,s,c,b,t$).  After renormalization,
these constants become scale dependent:
\begin{eqnarray}
g \rightarrow \alpha_s(\mu^2) &=&
\frac{12\pi}{33-2N_f}\,(\ln \mu^2/\Lambda^2)^{-1}\{1 + \ldots\}
\nonumber \\
m_a \rightarrow m_a(\mu^2) &=& \bar{m}_a \, (\frac{1}{2}\ln
\mu^2/\Lambda^2)^{-\frac{12}{33-2N_f}}\{ 1 + \ldots\}.
\end{eqnarray}
(The dots stand for higher loop corrections.)  Experimental study of
these parameters is necessarily indirect, since quarks and gluons do
not exist as free particles.  The measurement of the running coupling
constant $\alpha_s(\mu^2)$ in different processes and at different
scales provides a nontrivial test of QCD at short
distances.\cite{alphas} Heavy quark (c,b,t) masses can -- in principle
-- be investigated in the framework of heavy flavor quarkonia
spectroscopy.  Finally, as argued below, hadronic $\tau$ decays seem
to provide a unique source of experimental information on light
(current) quark running masses $m_u,m_d,m_s$.\cite{qm}  The purpose of the
present communication is to indicate how this information can be
extracted in practice from experiment.

Light quark masses (LQM) have a reputation of being "not
well-measurable"\cite{deruj} and, indeed, their experimental
determination has so far not even been attempted.  On the other hand,
there exists a huge number of theoretical
estimates\cite{qm,poubelle,narison,ddr,ms-mu} of their values, some of
which claim an amazing accuracy.\cite{ddr,ms-mu}  These estimates will
be commented on shortly.  At any rate, LQM are certainly tiny compared
to the scale $\Lambda_H ~ \sim $ 1\,GeV at which massive bound states
of QCD are formed. This fact represents the major difficulty in the
experimental study of LQM, since their contribution to hadron masses
(except pions) is small and hard to estimate theoretically.  On the
other hand, LQM measure the absolute
strength of chiral symmetry breaking.  This follows from the fact that
in QCD the divergences of observable axial and vector weak-transition
currents are given by the equations
\begin{mathletters}
\begin{eqnarray}
\partial^\mu(\bar{d} \gamma_\mu \gamma_5 u)
&=& (m_d + m_u) \bar{d}i \gamma_5 u \\
\partial^\mu(\bar{s} \gamma_\mu \gamma_5 u)
&=& (m_s + m_u) \bar{s}i \gamma_5 u \\
\partial^\mu(\bar{d} \gamma_\mu u)
&=& (m_d - m_u) \,i \bar{d} u \\
\partial^\mu(\bar{s} \gamma_\mu  u)
&=& (m_s - m_u) \,i \bar{s} u
\end{eqnarray}
\end{mathletters}
which are valid at all scales.  The experimental determination of LQM
is based on these equations, combined with the short distance
properties of QCD.

Before we develop the details of our argument that experimental
investigation of LQM is feasible, let us briefly summarize why it is
important:
\begin{itemize}
\item[(i)] It is important to know $all$ parameters of the standard model.
\item[(ii)] The values of LQM are closely related\cite{ddr} to the
value of the {\em quark-antiquark condensate} $\langle \bar{\psi}\psi
\rangle$, which is an important quantitative characteristic of the
nonperturbative chiral order in the QCD vacuum.
\item[(iii)] It is important to have a $direct$ experimental check of
the theoretical prediction for the {\em ratios of LQM} that is based on
standard chiral perturbation theory.\cite{gl85}  The existence of a
possible disagreement between this prediction and some low-energy data
has recently been pointed out.\cite{fss90,ssf93}
\item[(iv)] The claim that "$m_u$ is not equal to zero"\cite{leut}
should be framed in terms of a bound referred to a confidence level
with a firm experimental basis.  This issue is important for the
understanding of the strong CP violation problem.
\end{itemize}

{\flushleft{\bf 2.  HOW BIG IS THE DIVERGENCE OF THE AXIAL CURRENT?}}

We shall mainly concentrate on the average of the $u$ and $d$ quark
masses,
\begin{equation}
\hat{m} = (m_u + m_d)/2,
\end{equation}
which, according to Eq.\,(2a), controls the strength of the divergence
of the axial current.  (The remaining cases of $m_s - m_u$ and $m_d -
m_u$ will be briefly mentioned in Section 5.)  The object of our
concern will be the spectral function
\begin{equation} \rho (Q^2) =
\frac{1}{2\pi} \sum_n (2\pi)^4 \delta^{(4)}(Q - P_n) \,| \langle n |
\partial^\mu (\bar{d}\gamma_\mu \gamma_5 u) | 0 \rangle |^2,
\end{equation}
where the sum extends over all states with the quantum numbers of the
pion and with squared invariant mass $Q^2$:
\begin{equation}
n = \pi^- ,\, \pi^- \pi^+ \pi^- ,\, \pi^- \pi^0 \pi^0, \,
\pi^- \pi^+ \pi^- \pi^0 \pi^0 , \ldots .
\end{equation}
The spectral function $\rho(Q^2)$ measures the amount of explicit chiral
symmetry breaking at squared momentum transfer $Q^2$.  For large
$Q^2$, it is given \cite{ddr} by QCD perturbation theory:
\begin{equation}
\rho(Q^2) \rightarrow \frac{3}{2\pi^2} [\hat{m}(Q^2)]^2 \, Q^2 \left\{
1 + \frac{17}{3}\frac{\alpha_s(Q^2)}{\pi} + \ldots \right\}.
\end{equation}
Hence, $\hat{m}(Q^2)$ is $directly$ measurable to the extent that
$\rho(Q^2)$ is measurable for sufficiently large $Q^2$.  The spectral
function is comprised of individual exclusive components
\begin{equation}
\rho(Q^2) = 2 F_\pi^2 M_\pi^4 \delta(Q^2 - M_\pi^2) + \rho_{3\pi}(Q^2)
+ \rho_{K\bar{K}\pi}(Q^2) + \rho_{5\pi}(Q^2) + \ldots .
\end{equation}
The one-pion contribution is the only one which is known; it is given
by the pion mass, $M_\pi$ = 139 MeV, and by the pion decay constant,
$F_\pi$ = 93.1 MeV.  The remaining components,
$\rho_{3\pi},\rho_{K\bar{K}\pi}$, etc. are {\em terra incognita} of
particle physics; they hide the experimental information on the quark
mass $\hat{m}$  that we are looking for.

The only way to measure the unknown components of the spectral
function Eq.\,(4) seems to be $via$ {\em exclusive hadronic $\tau$
decays}, such as $\tau \rightarrow \nu_\tau + 3\pi, \nu_\tau + 5\pi,
\nu_\tau + \pi K\bar{K}$, etc. (See Fig.\,1.) In order to extract the
desired information from these decays, one faces two distinct problems:
\begin{itemize}
\item[(i)] The final hadronic state (say $3\pi$) excited from the
vacuum by the virtual $W$ consists of $J=1$ and $J=0$ waves.  The
$J=1$ wave, which is large and resonant (cf. the $a_1$ resonance), is
a subject of current experimental study \cite{a1}; however, for our
purposes, it is of little interest. $\rho_{3\pi}(Q^2)$ is given by the
square of the $J=0$ part of the corresponding amplitude, which is
$O(\hat{m}^2)$, i.e. too small to be directly measured.  A
model-independent solution of this problem is presented in the next
section.  $\rho_{3\pi}$ can be reconstructed from the {\em
interference of J=0 and J=1 waves}, which is $O(\hat{m})$, and
can be measured even for $\hat{m}$ as small as 7 MeV, provided one has
large enough statistics.
\item[(ii)] The second difficulty is related to the use of the
asymptotic formula Eq.\,(6) in a not quite asymptotic region of $Q^2$.
Not only is $Q^2$ limited by the $\tau$ mass $m_\tau$, but the
differential decay rate is strongly suppressed near $Q^2 = m_\tau^2$.
This difficulty may be resolved by using QCD sum rules\cite{svz} for
the two-point correlator of the axial current divergence.  In general,
these sum rules can be put into the form
\begin{equation}
\label{sumrules}
\hat{m}^2(s_0) = H^{-1}(w,s_0) \int_0^\infty dQ^2 \ w(Q^2,s_0) \
\rho(Q^2).
\end{equation}
Here, $w$ denotes a positive weight function that selects
contributions with $Q^2 < s_0$.  In practice, one uses $w =
exp(-Q^2/s_0)$ or $w = Q^{2n}\theta(s_0 - Q^2)$.  $H(w,s_0)$ is then
defined by the large $Q^2$ behavior of the two-point correlator.  It
consists of a part given by QCD  perturbation theory and of a
nonperturbative part parametrized in terms of vacuum condensates.  The
theoretical uncertainty in $H(w,s_0)$ decreases with increasing $s_0$.
\end{itemize}

Sum rules of the type shown in Eq.\,(\ref{sumrules}) have been
extensively used in the past
to estimate quark masses.  These studies show a reasonable stability
with respect to different choices of the weight function $w$, and they
are rather insensitive to the uncertainty in the nonperturbative part
of $H(w,s_0)$.  The main problem in these estimates is not the
sum-rule technique itself but rather the {\em complete absence of
experimental information on the magnitude of $\rho(Q^2)$} beyond the
one-pion contribution.  Retaining just the pion contribution to the
integral of Eq.\,(\ref{sumrules}), and using the positivity of individual
components Eq.\,(7) of $\rho(Q^2)$, one finds\cite{narison} a {\em
lower bound}
\begin{equation}
\hat{m}({\rm 1~GeV^2}) \ge {\rm (4-5)~MeV.}
\end{equation}
We propose to exploit the sum rules Eq.\,(\ref{sumrules}) once more,
and to improve
the lower bound Eq.\,(9) by using the $measured$ component
$\rho_{3\pi}(Q^2)$ as input. $\tau$ decays make it possible to work at
higher $s_0$ ($s_0 \le m_\tau^2$), thus reducing the systematic error
due to the nonperturbative part of  $H(w,s_0)$.

{\flushleft {\bf 3. RECONSTRUCTION OF $\rho_{3\pi}$ FROM $\tau$-DECAY
EXPERIMENTS}}

There are two $3\pi$ contributions to $\rho(Q^2)$:  $\pi^- \pi^- \pi^+$
and $\pi^0 \pi^0 \pi^-$.  They are not related by any symmetry, and
consequently they should be measured separately and added afterwards:
\begin{equation}
\rho_{3\pi}(Q^2) = \rho_{_{--+}}(Q^2) + \rho_{_{00-}}(Q^2).
\end{equation}
The kinematic analysis of the corresponding $\tau^- \rightarrow
\nu_\tau 3\pi$ decays is identical in both cases; the difference
resides merely in possible experimental difficulty in detecting
neutral pions with sufficient efficiency.  Even if the latter could
not be attained, experimental information on $\rho_{_{--+}}$ would
still be valuable.  This is because isospin symmetry implies the
inequality
\begin{equation}
\rho_{3\pi}(Q^2) \ge \frac{5}{4} \rho_{_{--+}}(Q^2),
\end{equation}
which could lead to a considerable improvement of the lower bound Eq.\,(9)
even if no experimental information on $\rho_{_{00-}}$ were available.

{\flushleft {\bf 3a. Azimuthal asymmetries}}

The kinematics of exclusive decays of $\tau$ into three hadrons has
been analyzed in full generality in Ref.\cite{km} (hereafter referred
to as KM).  Here, we will use the same notation, except for a few
simplifications which are appropriate for the special case of the
decays
\begin{equation}
\tau^- \rightarrow \pi^a(q_1) \pi^a(q_2) \pi^b(q_3) \nu_\tau,
\end{equation}
where $a=-(0),~b=+(-)$.  These decays proceed $via$ the axial weak
current
\begin{equation}
\langle \pi^-(q_1) \pi^-(q_2) \pi^+(q_3) | \bar{d}\gamma_\mu \gamma_5
u | 0 \rangle \equiv A_\mu(q_1,q_2,q_3)
\end{equation}
and are described by three independent form factors.  (The anomalous
vector current contribution vanishes because of G-parity
conservation.)  Following KM, it is convenient to work in the hadronic
center of mass frame, in which the three pions move in the $x-y$
plane.  The $x$ axis is chosen parallel to $\vec{q}_3$ and the
hadronic plane is oriented so that the $z$ axis points in the
direction of $\vec{q}_1\times\vec{q}_2$, where $|\vec{q}_1| >
|\vec{q}_2|$.  In this frame, the $z$-component of Eq.\,(13) vanishes
and the three independent components $A_x, A_y$ and $A_t$ are
functions of the scalar hadronic variables
\begin{equation}
Q^2 = (q_1 + q_2 + q_3)^2,~s_1 = (Q-q_1)^2,~ s_2 = (Q-q_2)^2.
\end{equation}
The three form factors $A_x, A_y$ and $A_t$ are related to the form
factors $F_1, F_2$ and $F_4$ of KM by
\begin{equation}
A_x = x_1 F_1 + x_2 F_2,~ A_y = x_3(F_1 - F_2),~A_t = \sqrt{Q^2}F_4,
\end{equation}
where the $x_i$ are kinematic functions defined in Eq.\,(33) of KM.
The space components $A_x$ and $A_y$ describe the $J=1$ part of
the matrix element.  They are large and insensitive to the quark mass.
The time component $A_t$ is proportional to the matrix element of the
divergence of the axial current, and consequently to the running quark
mass $\hat{m}$.

We do not assume a polarized $\tau$ beam, and the knowledge of the
direction of flight of the $\tau$ in the lab system (i.e., of the
$\tau$ rest frame) is not needed.  One should just measure the momenta
of the three pions and reconstruct their center of mass frame as
defined above.  This determines $Q^2$, the Dalitz plot variables $s_1$
and $s_2$, and the angles $\beta$ (polar) and $\gamma$ (azimuthal)
which define the direction of flight $\vec{n}_L$ of the laboratory
with respect to the oriented center of mass hadronic plane (see
Fig.\,2). The remaining variable is then $\theta$ (the $\tau$
decay angle) given in terms of the hadronic energy $E_h$ and the
energy $E_\tau$ of the $\tau$, both in the laboratory frame.

The hadronic structure functions of interest are
\begin{equation}
W_{ab}(Q^2,\Delta_i) \equiv \int_{\Delta_i} ds_1 ds_2 Re(A_a^* A_b), ~~~
a,b = x,y,t,
\end{equation}
where $\Delta_i$ denotes a bin of the Dalitz plot.  $W_{xx}, W_{yy}$
and $W_{xy}$ are independent of the quark mass, while the interference
terms $W_{xt}$ and $W_{yt}$ are $O(\hat{m})$, and $W_{tt}$ is
$O(\hat{m}^2)$. The latter determines the contribution to the spectral
function Eq.\,(4),
\begin{equation}
\rho_{_{--+}}(Q^2) = \frac{1}{512\pi^4} \sum_i W_{tt}(Q^2,\Delta_i),
\end{equation}
where the sum extends over all bins of the Dalitz plot.  (By definition,
the sum in (17) is independent of the manner in which the phase space is
divided
into bins.) The most straightforward measurement of
the relevant structure functions $W_{ab}(Q^2,\Delta_i)$ involves the
differential decay rate integrated over the polar angle $\beta$
and over the $\tau$-decay angle $\theta$:
\begin{equation}
\Gamma(Q^2,\Delta_i,\gamma) = \frac{1}{(2\pi)^5}
\frac{G_F^2}{128m_\tau} |V_{ud}|^2 \biggl( \frac{m_\tau^2 - Q^2}{Q^2}
\biggr)^{\!\!2} \, \frac{m_\tau^2 + 2Q^2}{3m_\tau^2} W(Q^2,\Delta_i)
f(\gamma) \frac{d\gamma}{2\pi} dQ^2,
\end{equation}
where
\begin{equation}
W(Q^2,\Delta_i) \equiv W_{xx}(Q^2,\Delta_i) + W_{yy}(Q^2,\Delta_i) +
\frac{3m_\tau^2}{m_\tau^2 + 2Q^2} W_{tt}(Q^2,\Delta_i)
\end{equation}
describes the differential decay rate integrated over all angles.  The
normalized distribution in the azimuthal angle $\gamma$ is of the form
\begin{equation}
f(\gamma) = 1 + \lambda_2(A \cos 2\gamma + B \sin 2\gamma ) +
\lambda_1 (C_{LR} \cos \gamma + C_{UD} \sin \gamma ).
\end{equation}
The coefficients $A$ and $B$ are ``large,'' i.e., they are $O(1)$ in
the chiral limit:
\begin{equation}
A = \frac{m_\tau^2 - Q^2}{m_\tau^2 + Q^2}\, \frac{W_{xx} -
W_{yy}}{W},~~~~
B = \frac{m_\tau^2 - Q^2}{m_\tau^2 + Q^2}\, \frac{2W_{xy}}{W}.
\end{equation}
The small chiral symmetry breaking shows up through the left-right and
up-down asymmetry coefficients
\begin{equation}
C_{LR} = \frac{\pi}{2} \frac{3m_\tau^2}{m_\tau^2 + 2Q^2}
\frac{W_{xt}}{W},~~~~
C_{UD} = -\frac{\pi}{2} \frac{3m_\tau^2}{m_\tau^2 + 2Q^2}
\frac{W_{yt}}{W},
\end{equation}
which are proportional to the quark mass $\hat{m}$.  Measurement of
these azimuthal asymmetries represents the hard core of our method of
determining $\hat{m}$.  The coefficients $\lambda_1$ and $\lambda_2$
in Eq.\,(20) are kinematic functions of $Q^2$ and of the velocity of
the $\tau$ in the laboratory frame, $\beta_\tau =
\sqrt{1-m_\tau^2/E_0^2}$, where $E_0$ is the energy of the beam.  The
$\lambda_n$ result from integration over the $\tau$ decay angle
$\theta$,
\begin{equation}
\lambda_n(Q^2,\beta_\tau) = \int_{-1}^1 \frac{d\cos \theta}{2}
P_n(\cos \psi),
\end{equation}
where $P_n$ are Legendre polynomials and $\cos \psi$ is a function of
$\cos \theta, Q^2$ and $\beta_\tau$ as defined in KM.  The shape
of these functions depends strongly upon the energy of the machine,
$E_0$.  For $\tau$ leptons produced at rest $\cos \psi = 1$, so one
has
\begin{mathletters}
\begin{equation}
\lambda_n(Q^2,0) \equiv 1,
\end{equation}
whereas in the ultrarelativistic limit $\beta_\tau = 1$ these
functions become (see Fig.\,3.)
\begin{eqnarray}
\lambda_1(Q^2,1) &=& \frac{m_\tau^4 - Q^4 + 2m_\tau^2 Q^2
\ln Q^2/m_\tau^2}{(m_\tau^2 - Q^2)^2}\nonumber \\
\lambda_2(Q^2,1) &=& -2 + 3\frac{m_\tau^2 + Q^2}{m_\tau^2 - Q^2}
\lambda_1(Q^2,1).
\end{eqnarray}
\end{mathletters}

For low-energy machines such as tau-charm factories, the integration
over the decay angle $\theta$ represents a simplification which
does not lower the sensitivity to the azimuthal asymmetries (21) and
(22).  On the other hand, for high-energy machines such as at LEP
($\beta_\tau \approx$ = 0.999), CESR ($\beta_\tau \approx$ = 0.93) or
B-factories,
Eq.\,(24b) indicates an important loss in sensitivity.  In this case,
one should make use of the knowledge of the decay angle distribution
rather than simply integrating over it.

{\flushleft{\bf 3b.  The trick}}

One may take advantage of the smallness of the quark mass and neglect
$W_{tt}$ compared to $W_{xx}+W_{yy}$ in Eq.\,(19).  Within this
approximation, the measurement described above yields the structure
functions $W_{xx},W_{yy},W_{xy}$ and $W_{xt},W_{yt}$ for a given $Q^2$
and for a given bin $\Delta_i$ of the Dalitz plot.  On the other hand,
similar information on the interference terms $Im\,A_x^*A_t$ and
$Im\,A_y^*A_t$ would require $both$ a known nonzero $\tau$
polarization $and$ the reconstruction of the $\tau$ rest frame.  This
information is, fortunately, not needed.  We will prove that, provided
the binning of the Dalitz plot is sufficiently fine, the
$O(\hat{m}^2)$ quantity $W_{tt}$ can be reconstructed from the
experimentally accessible $O(\hat{m})$ interference terms
\begin{mathletters}
\begin{equation}
\phi(Q^2,\Delta_i) = \left (
\begin{array}{c}
W_{xt}\\
W_{yt}
\end{array}
\right )
\end{equation}
and from the spin-1 structure functions
\begin{equation}
{\bf\sf K}(Q^2,\Delta_i) = \left (
\begin{array}{cc}
W_{xx} & W_{xy}\\
W_{yx} & W_{yy}
\end{array}
\right ) = {\bf\sf K}^T.
\end{equation}
\end{mathletters}

Given two complex numbers $x$ and $z$, it is obviously impossible to
reconstruct $|z|^2$ from $|x|^2$ and $Re(x^*z)$ alone.  However, for
$three$ complex numbers $x_1,x_2$ and $z$ one has the identity
\begin{equation}
|z|^2 = \sum_{i,j=1}^{2} Re(x_i^* z) ({\bf\sf k}^{-1})_{ij} Re(z^* x_j),
\end{equation}
where ${\bf\sf k}^{-1}$ denotes the inverse of the matrix ${\bf\sf
k}_{ij} =  Re(x_i x_j^*)$.  This identity applies to the decay $\tau
\rightarrow 3\pi \nu_\tau$, which is characterized by three (complex)
form factors.  For each bin $\Delta_i$, let us define the quantity
\begin{equation}
\overline{W}_{tt} (Q^2,\Delta_i) \equiv \phi^T(Q^2,\Delta_i)
{\bf\sf K}^{-1}(Q^2,\Delta_i) \phi(Q^2,\Delta_i)
\end{equation}
given entirely in terms of observables (25a) and (25b).  If a bin
$\Delta$ shrinks to a point $P$ of the Dalitz plot, one has
\begin{equation}
W_{ab}(Q^2,\Delta) = Re(A_a^*(P) A_b(P))~\epsilon(\Delta)
\end{equation}
where
\begin{equation}
\epsilon(\Delta) = \int_\Delta ds_1 ds_2 \rightarrow 0.
\end{equation}
Consequently, in the small bin limit, the indentity (26) implies
\begin{equation}
W_{tt}(Q^2,\Delta) = \overline{W}_{tt}(Q^2,\Delta) + O(\epsilon^2).
\end{equation}

It is now straightforward to reexpress this result in terms of the
spectral function Eq.\,(17).  For a given division of the Dalitz plot
into N bins, define
\begin{equation}
\rho_{_{--+}}^{(N)}(Q^2) \equiv \frac{1}{512\pi^4} \sum_{i=1}^N
\overline{W}_{tt}(Q^2,\Delta_i),
\end{equation}
where $\overline{W}_{tt}$ is defined by Eq.\,(27).  Consider now the
limit $N \rightarrow \infty$ such that $\epsilon(\Delta) \sim
O(N^{-2})$.  The "true" spectral function is then obtained as
\begin{equation}
\rho_{_{--+}}(Q^2) = \lim_{N\rightarrow\infty}
\rho_{_{--+}}^{(N)}(Q^2).
\end{equation}

In practice, however, one deals with a finite number of bins of a
finite size, and it is important to analyze what can be concluded in
this case. Independently of the size of $\Delta$, Eq.\,(16) defines a
scalar product which satisfies the Schwarz inequality
\begin{equation}
W_{tt}(Q^2,\Delta) \ge
\frac{|W_{xt}(Q^2,\Delta)|^2}{W_{xx}(Q^2,\Delta)}.
\end{equation}
A similar lower bound holds under the substitutions $W_{xt}
\rightarrow W_{xt}'$ and $W_{xx} \rightarrow W_{xx}'$, where $W_{xt}'$
and $W_{xx}'$ are obtained by the replacement
\begin{equation}
A_x \rightarrow A_x' = g_1A_x + g_2A_y,
\end{equation}
$g_1,g_2$ being two arbitrary $real$ constants.  In this way, one
generates a whole class of lower bounds, and one can then ask which
one is the best. Remarkably, the answer leads to the reappearance of
the expression Eq.\,(27); one easily finds
\begin{equation}
\max_{(g_1,g_2)} \frac{(W_{xt}')^2}{W_{xx}'} =
\overline{W}_{tt}(Q^2,\Delta_i).
\end{equation}
Consequently, for any finite size bin $\Delta_i$, one has the {\em
lower bound}
\begin{equation}
W_{tt}(Q^2,\Delta_i) \ge \overline{W}_{tt}(Q^2,\Delta_i),
\end{equation}
which, according to Eq.\,(30), is saturated in the limit of small
bins.  For $any$ splitting of the Dalitz plot into $N$ bins, one can
sum Eq.\,(36) over all bins and divide by $512\pi^4$; by Eq.\,(31),
the left hand side is simply $\rho_{_{--+}}(Q^2)$, which is, of
course, independent of binning, while the right hand side is precisely
$\rho_{_{--+}}^{(N)}(Q^2)$.  One thus obtains the lower bound
\begin{equation}
\rho_{_{--+}}(Q^2) \ge  \rho_{_{--+}}^{(N)}(Q^2),
\end{equation}
which is saturated with increasing number of bins.

It is worth noting the independence of our method on the geometry
chosen in the hadronic plane.  This is a reflection of the fact that
the expression Eq.\,(27) is invariant under
\begin{equation}
\phi \rightarrow g\phi,~~~~~{\bf\sf K} \rightarrow g{\bf\sf K}g^T,
\end{equation}
where $g$ is a two-by-two real matrix, representing a general linear
(nonsingular) $real$ transformation of a two-dimensional space spanned
by $A_x$ and $A_y$.

{\flushleft{\bf 3c.  Back to the sum rules}}

Suppose that the above program has been completed for both $\tau^-
\rightarrow \pi^-\pi^-\pi^+\nu_\tau$ and $\tau^- \rightarrow
\pi^0\pi^0\pi^-\nu_\tau$ decays.  That is, the azimuthal asymmetries
have been measured in $N$ bins of the Dalitz plot and
the corresponding functions $\rho_{_{--+}}$ and $\rho_{_{00-}}$ have
been constructed according to Eq.\,(31) for $N\le N_{max}$ allowed by
the available statistics.  One can then return to the QCD sum rules
(\ref{sumrules}) and define the quantity
\begin{eqnarray}
\hat{m}^2_N (\mu,s_0) \equiv \biggl( \frac{\ln s_0/\Lambda^2}{\ln
\mu^2/\Lambda^2} \biggr) ^{24/29}  H^{-1}(w,s_0) \int_0 ^{m_\tau^2}
dQ^2 w(Q^2,s_0) \{ 2F_\pi^2 M_\pi^4 \delta(Q^2 - M_\pi^2) \nonumber \\
+ \rho_{_{--+}}^{(N)}(Q^2) + \rho_{_{00-}}^{(N)}(Q^2) \}.
\end{eqnarray}
One expects that $\hat{m}_N(\mu,s_0)$ will depend rather weakly on the
choice of the weight function $w$, and on $s_0$ in a typical range
\begin{equation}
{\rm 2~GeV^2} \le s_0 \le m_\tau^2 \simeq {\rm 3.18~GeV^2},
\end{equation}
especially for large $N$.  For $s_0$ in the range (40), the running
quark mass $\hat{m}(\mu)$ obeys
\begin{equation}
\hat{m}(\mu) \ge \hat{m}_N(\mu,s_0),
\end{equation}
and this lower bound should saturate rather rapidly with increasing
$N$.  The saturation as well as the variation of the right hand side
of Eq.\,(39) with $s_0$ can be controlled experimentally.  The latter
variation can be considered as a source of systematic error arising
from imperfections in the method.  Another source of error, which is
more difficult to estimate, is related to both perturbative (higher
orders \cite{3loop}) and nonperturbative (condensates,
instantons \cite{instantons}) uncertainties in the high-energy factor
$H(w,s_0)$.

{\flushleft{\bf 4.  WHAT IS EXPECTED}}

Light quark masses are the only entries in the Particle Data Group
compilations \cite{pdg} that are not based on a measurement but on
theoretical estimates.  We first briefly recall why a direct estimate
of $\hat{m}$ is problematic without experimental information on
$\rho_{3\pi}(Q^2)$. Then we proceed to a model-dependent numerical
study of the various steps of  the method described in the preceding
section.  In particular, the resulting statistical error of a
measurement of $\hat{m}$(1 GeV) will be estimated, using the maximum
likelihood method.

For definiteness, the finite energy version of QCD sum rules will be
used here, closely following Ref.\cite{ddr}; Eq.\,(\ref{sumrules})
then takes the form
\begin{equation}
\hat{m}^2(s_0) = \frac{4\pi^2}{3s_0^2} \left[ 1 + R_2(s_0) +
2 C_4 \langle O_4 \rangle /s_0^2 \right] ^{-1} \int_0^{s_0} dQ^2
\rho(Q^2),
\end{equation}
and similarly, the definition Eq.\,(39) becomes
\begin{eqnarray}
\label{fesr}
\hat{m}^2_N (\mu,s_0) \equiv \biggl( \frac{\ln s_0/\Lambda^2}{\ln
\mu^2/\Lambda^2} \biggr) ^{\!24/29} \,\frac{4\pi^2}{3s_0^2}   \left[ 1 +
R_2(s_0
) +
2 C_4 \langle O_4 \rangle /s_0^2 \right] ^{-1} \times \nonumber \\
\left[ 2F_\pi^2 M_\pi^4 + \int_0 ^{s_0}
dQ^2 \{ \rho_{_{--+}}^{(N)}(Q^2) + \rho_{_{00-}}^{(N)}(Q^2) \} \right],
\end{eqnarray}
where the two-loop expression for $R_2(s_0)$ as well as a discussion
of the value of the dimension-4 condensate $C_4 \langle O_4 \rangle$
can be found in Ref.\cite{ddr}.  We will use the value for the QCD
coupling constant $\alpha_s$ as determined in Ref.\cite{alphas}.

{\flushleft{\bf 4a. Chiral Perturbation theory}}

Unlike $\hat{m}$, the order of magnitude of the difference $m_s -
\hat{m}$ is fairly easy to estimate:  From the size of $SU_V(3)$
symmetry breaking, one may infer
\begin{equation}
\label{msmhat}
(m_s - \hat{m})({\rm 1~GeV^2}) = 100 - 300~ {\rm MeV},
\end{equation}
in agreement with sum-rule results.\cite{ms-mu}  (A possible
measurement of $m_s - m_u$ in $\tau \rightarrow \nu_\tau K\pi$ decay
will be briefly mentioned in the following section.)  The estimate
Eq.\,(\ref{msmhat}) can be combined with the $presumed$ value of the quark
mass ratio $r = m_s/\hat{m}$ = 26 to conclude that $\hat{m}({\rm
1~GeV^2}) = 4 - 12~ {\rm MeV}$, in agreement with the lower bound (9).
Actually, even this rather conservative and crude estimate of
$\hat{m}$ is doubtful.  The value $r$ = 26 is based on the standard
chiral perturbation theory which assumes that the quark antiquark
condensate $-\langle \bar{\psi}\psi \rangle$ is large compared to
$F_\pi^2 m_s$.  The latter hypothesis has no clear experimental or
theoretical support, and in fact it need not be correct \cite{fss90,ssf93}
for the actual value of $m_s$.  The generalized chiral perturbation
theory \cite{ssf93} which admits a lower value of $-\langle \bar{\psi}\psi
\rangle$ does not fix the quark mass ratio $r = m_s/\hat{m}$; any
value $6.3 \le r \le 25.9$ is consistent with the mass spectrum of
pseudoscalar mesons, the lower bound for $r$ arising from the
condition of vacuum stability.  Hence, accepting the estimate
Eq.\,(\ref{msmhat}) constrains $\hat{m}$ to the range
\begin{equation}
{\rm 4~ MeV} \le \hat{m}({\rm 1~GeV^2}) \le {\rm 50~ MeV}.
\end{equation}
However, it is clear that finding $\hat{m}$ significantly higher than
-- say -- 10 MeV would imply a considerably lower value of $-\langle
\bar{\psi}\psi \rangle$ and of $r = m_s/\hat{m}$ than the standard
chiral perturbation theory could support.  (Independent, though
indirect, measurements of the quark mass ratio $r$ are possible in
low-energy $\pi-\pi$ scattering,\cite{ssf93}
$K_{\mu4}$ decays \cite{knecht} and from observed
corrections to the Goldberger-Treiman relations.\cite{fss90})  This
illustrates once more why even an experimental {\em lower bound} on
$\hat{m}$ would be of considerable interest.

It is instructive to illustrate the variation of $\hat{m}$ and
$\langle \bar{\psi}\psi \rangle$ within a simple model, in which the
$3\pi$ contribution to the spectral function is described by a narrow
$J^P = 0^-$ resonance -- the $\pi'$:
\begin{equation}
\label{model}
\rho_{3\pi}(Q^2) = 2F_{\pi '}^2 M_{\pi '}^4 \,\delta(Q^2 - M_{\pi '}^2).
\end{equation}
(Such a resonance does indeed exist for $M_{\pi '} \approx$ 1.3 GeV,
but it is not narrow.\cite{pdg})  The constant $F_{\pi '}$ -- the $\pi
'$ analogue of $F_\pi$ = 93.1 MeV -- describes the coupling of the
$\pi '$ to the axial current.  It is proportional to $\hat{m}$
($F_{\pi '}$ = 0 in the chiral limit) and it is expected to be small
compared to $F_\pi$.  Its value is, however, unknown and there is no
reliable model available to pin it down.  (In particular, quark model
estimates (e.g., Ref.\cite{isgur}) of $F_{\pi '}$ are trustworthy only
to the extent that the
model would correctly describe the small breaking of chiral symmetry.)
Considering $F_{\pi '}$ as a free parameter, one may use the narrow
$\pi '$ model Eq.\,(\ref{model}) within the FESR Eq.\,(42) in
order to investigate the sensitivity of $\hat{m}$ to $F_{\pi '}$.  Setting
$M_{\pi '}$ = 1.3 GeV and fixing $s_0$ from the higher moment sum
rules (as explained in Ref.\cite{ddr}), one finds that as $F_{\pi '}$
increases from 1.5 MeV to 15 MeV, $\hat{m}$(1 GeV) slowly rises from 7
MeV to 45 MeV.  At the same time, the value of the $\bar{q}q$
condensate $-F_\pi^{-2} \langle \bar{\psi}\psi \rangle$ rapidly decreases
from $\sim$ 1.3 GeV to $\sim$ 40 MeV.  {\em A priori}, there is no reason to
exclude a value for $F_{\pi '}$ as large as 10 or 15 MeV.  However, it
is clear that, in this case, the $\pi '$ (or $3 \pi$) contribution to
the quark mass would largely dominate over the single pion
contribution.  Moreover, the slow rise of $\hat{m}$ would not
compensate for the drop of $-\langle \bar{\psi}\psi \rangle$, so that
$-2\hat{m}\langle \bar{\psi}\psi \rangle$  would be considerably lower
than $F_\pi^2M_\pi^2$.  Such a violation of the Gell-Mann, Oakes,
Renner formula is allowed and expected within the generalized chiral
perturbation theory,\cite{ssf93} and its experimental confirmation
would be an argument in favor of the latter.

It is conceivable that the $\pi '$ width cannot be neglected, and that
a more realistic model of $\rho_{3\pi}(Q^2)$ is provided by a
Breit-Wigner formula.  However, even if the corresponding width were
known, the absolute normalization of $\rho_{3\pi}(Q^2)$ remains
unspecified.  Dominguez and de Rafael \cite{ddr} have proposed to
normalize $\rho_{3\pi}(Q^2)$ by its {\em low $Q^2$ behavior}
\begin{equation}
\label{rho3pi}
\rho_{3\pi}(Q^2) \rightarrow \frac{1}{768\pi^4}
\frac{M_\pi^4}{F_\pi^2} Q^2,
\end{equation}
as given by chiral perturbation theory. In this way, they obtain
$\hat{m}$(1 GeV) = 7.8$\pm$1.0 MeV, where the quoted error merely
arises from the input data ($M_{\pi '},\Gamma_{\pi '}$) and from the
uncertainty brought in by the sum rules.  It does not involve the
possible error in the absolute normalization of  $\rho_{3\pi}(Q^2)$
based on Eq.\,(\ref{rho3pi}).

Apart from doubts about using the $Q^2 \sim 0$ behavior of
$\rho_{3\pi}$ to fix its value at $Q^2 \sim M_{\pi '}^2 \sim {\rm
1.69~GeV^2}$, the main uncertainty in the above method of normalizing
$\rho_{3\pi}$ resides in the formula Eq.\,(\ref{rho3pi}) itself.  This
formula is obtained using the {\em standard} chiral perturbation
theory in an experimentally unexplored domain of low-energy $\pi-\pi$
interaction, where results strongly depend on the value of the quark
antiquark condensate $\langle \bar{\psi}\psi \rangle$.  The {\em
generalized} chiral perturbation theory,\cite{ssf93} which
parametrizes this dependence in a model independent way, leads to a
modification of Eq.\,(\ref{rho3pi}):
\begin{equation}
\rho_{3\pi}(Q^2) \rightarrow \frac{1}{768\pi^4}
\frac{M_\pi^4}{F_\pi^2} \, \frac{5\alpha_{\pi\pi}^2 + 1}{6} Q^2.
\end{equation}
Here, $\alpha_{\pi\pi}$ is a parameter introduced in \cite{ssf93}
which is a function of the quark mass ratio $r = m_s/\hat{m}$, varying
from $\alpha_{\pi\pi}$ = 1 ($r$ = 25.9) to $\alpha_{\pi\pi}$ = 4 ($r$
= 6.3), and which can be measured in low-energy $\pi-\pi$
scattering.\cite{ssf93} Hence, $\rho_{3\pi}(Q^2)$ cannot be normalized
using its low-$Q^2$ behavior, since the latter is only known up to a
factor of 1--13.5.  $\rho_{3\pi}(Q^2)$ has to be determined
experimentally.

{\flushleft{\bf 4b.  A model-dependent theoretical experiment}}

In order to show how the method above might work in practice, we have
generated data using a model for the form factors of Eq.\,(15)
and performed the analysis including an estimate of statistical errors.
We do not expect any model to give better than an order of magnitude
estimate of the observable quantities that we seek.
The form factors $F_1,F_2$ are chosen to be exactly those of
KM (cf. \cite{isgur,feindt}).  For the $J=0$ form factor $F_4$ we
follow KM in assuming the dominance of the $\pi '$ resonance that
decays into $\rho\pi$; however, we use the minimal $\pi ' \pi\rho$
coupling (which they do not), so that
\begin{equation}
F_4(s_1,s_2,Q^2) = -35i \, \xi \,
BW_{\pi '}(Q^2) [(s_2 - s_3)B_\rho(s_1) + (s_1 - s_3)B_\rho(s_2) ]
\end{equation}
where $s_3 = Q^2 + 3M_\pi^2 -s_1 -s_2$ and the $\pi '$ Breit-Wigner
is given by
\begin{eqnarray}
BW_{\pi '}(Q^2) &=& \frac{M_{\pi '}^2}
{M_{\pi '}^2 - Q^2 -i\sqrt{Q^2}\Gamma_{\pi '}(Q^2)}\nonumber \\
\Gamma_{\pi '}(Q^2) &=& \Gamma_{\pi '} \frac{M_{\pi '}^2}{Q^2}
[ p(Q^2)/p(M_{\pi '}^2)] ^3 \nonumber \\
p(Q^2) &=& \sqrt{[Q^2 - (M_\rho + M_\pi)^2][Q^2 - (M_\rho -
M_\pi)^2]/(4 Q^2)}.
\end{eqnarray}
The parameters $M_{\pi '}$ and $\Gamma_{\pi '}$ as well as
the spin-1 function, $B_\rho$, are taken directly from KM without
alteration.  For simplicity, we take the pion mass to be zero.  We have
introduced a dimensionless parameter $\xi$ which
sets the scale of the $J=0$ form factor (the numerical factor of 35 is
for convenience, with units $\rm GeV^{-3}$ understood here).
The parameter $\xi$ plays the same role here as
the constant $F_{\pi '}$ did in the narrow-resonance model described
in the preceding subsection; that is, it is
$\xi$ which determines the contribution of the $J=0$ spectral function
to the quark mass, and so it is an unknown quantity.  We will see below
that our normalization is chosen so that, within the framework of the
model we have adopted, $\xi$ is of order unity for values of $\hat{m}$
in the range (45). Note that the model treats the
$\pi^-\pi^-\pi^+$ and $\pi^0\pi^0\pi^-$ modes similarly, so we may
simply combine the effects of the two modes by absorbing all
normalization into the single parameter $\xi$.

The first step in our experiment is to generate the data, i.e.,
compute the differential decay rate $\Gamma(Q^2,\Delta_i,\gamma)$ of
Eq.\,(18).  We do this for three cases, $\xi$ = 0.5, 1 and 2.  Using
the form factors $F_1,F_2,F_4$ described above, we compute from
Eq.\,(14) the functions $A_x,A_y,A_t$; these are then used to
determine from Eq.\,(16) the functions $W_{ab}$, which then give
$\Gamma$.  For our three choices of $\xi$, we show in Fig.\,4 the
asymmetry coefficients $A,B,C_{LR},C_{UD}$ of Eqs.\,(21),(22), where
the $W_{ab}$ are integrated over the entire Dalitz plot.  Note that
the azimuthal asymmetry coefficients $C_{LR},C_{UD}$ are on the order
of a few percent or less, much smaller than $A$ ($B$ is accidentally
small here, which is due to the choice of axes in the hadronic plane).
This is expected, since $C_{LR},C_{UD}$ are $O(\hat{m})$, while $A,B$
are $O(1)$ in the chiral limit.  Similarly, we may compute the
functions $W_{xx},W_{yy},W_{xy},W_{xt},W_{yt}$ as functions of
$Q^2,s_1,s_2$.  In Fig.\,5 we show (for $\xi$ = 1) the functions
$W_{ab}$ integrated over the entire Dalitz plot; clearly $W_{tt}$ is
negligible.

Next, we study the lower bound of Eq.\,(43) for four choices of
binning of the Dalitz plot: divide the square $0 \le s_1,s_2 \le
m_\tau^2$ into 1, 4, 16, or 64 equal-sized squares; this gives a total
of $N$ = 1, 3, 10 and 36 bins respectively for these four cases, since
the region of the Dalitz plot is simply a triangular half of this
square region.  For a given choice of binning, we compute for each bin
$\Delta_i$ the lower bound $\overline{W}_{tt}(Q^2,\Delta_i)$ from
Eq.\,(27).  After summing over bins, we show (for $\xi$ = 1) in Fig.\,6
how $\sum_{i=1}^{N}\overline{W}_{tt}(Q^2,\Delta_i)$ approaches
$W_{tt}(Q^2)$ as the number of bins $N$ grows.  Then, we obtain
$\rho_{_{--+}}^{(N)}(Q^2)$ from Eq.\,(31), and thus determine
$\hat{m}_N (\mu,s_0)$ from Eq.\,(43). The latter is shown (for
$\xi$ = 1) in Fig.\,7b as a function of $s_0$ for various $N$.  (We
take $\mu$ = 1 GeV.)  Note that the bound is essentially independent
of $s_0$ for ${\rm 2 ~GeV^2} \lesssim s_0 \lesssim m_\tau^2 \approx
{\rm 3.18~ GeV^2}$.
The figure clearly indicates that the choice of $\xi=1$ corresponds to
a lower bound on $\hat{m}(\mu)$ of about 14 MeV.  Similar results are
presented in Fig.\,7a for $\xi = 0.5$ (a lower bound of 7 MeV) and in
Fig.\,7c for $\xi = 2$ (a lower bound of 28 MeV).  We see that the
bound is roughly proportional to $\xi$ in this range of $\xi$.  For
smaller $\xi$ the pion contribution to the spectral function
eventually dominates, and the bound on $\hat{m}(\mu)$ reduces to that
of Ref.\cite{narison}, independent of $\xi$.

{\flushleft{\bf 4c.  Maximum likelihood estimate of error}}

Apart from any systematic experimental error, there will be some error
arising from finite statistics; this can be estimated by the maximum
likelihood method.  For $N_{evt}$ events, the likelihood function is
\begin{equation}
{\cal L}(\hat{m}_N(\mu,s_0)) = \prod_{i=1}^{N_{evt}} \Gamma(X_i;
\hat{m}_N(\mu,s_0)),
\end{equation}
where $X_i$ denotes the measured values of the phase space variables
$Q^2,s_1,s_2,\gamma$ for the event $i$, and $\Gamma$ is normalized to
unity.  For fixed $\mu$ = 1 GeV and fixed $s_0$, there is a one-to-one
correspondence between the lower bound $\hat{m}_N(\mu,s_0)$ and $\xi$;
therefore, the measurement of $\xi$ is equivalent to the measurement
of $\hat{m}_N(\mu,s_0)$. The best estimate for $\xi$ is the value
which maximizes the likelihood $\cal L$, or equivalently $\ln {\cal
L}$; the estimated standard deviation for the measurement of $\xi$ is
simply
\begin{equation}
\sigma_{\xi} = \left[ N_{evt} \int \frac{1}{\Gamma}
\biggl(\frac{\partial\Gamma}{\partial\xi}\biggr)^2 \right]^{-\frac{1}{2}}.
\end{equation}
Hence, the standard deviation for the measurement of the lower bound
$\hat{m} \equiv \hat{m}_N(\mu,s_0)$ is
\begin{mathletters}
\begin{equation}
\sigma_{\hat{m}} = \sigma_{\xi} (d\hat{m}/d\xi);
\end{equation}
then, using Eq.\,(43) to compute $(d\hat{m}/d\xi)$ [the $\xi$
dependence of $\hat{m}$ is contained in $\rho_{++-}^{(N)}(Q^2)$ and
$\rho_{00-}^{(N)}(Q^2)$],
\begin{equation}
\sigma_{\hat{m}} = \sigma_{\xi} (\hat{m}/\xi)
[1 - \hat{m}^2_0/\hat{m}^2 ],
\end{equation}
\end{mathletters}
where $\hat{m}_0$ is obtained from $\hat{m}$ by putting
$\xi$ = 0, i.e., including only the contribution of the pion to the
integral of Eq.\,(43).  For our model calculation, the standard
deviation $\sigma_{\hat{m}}$ varies as $1/\sqrt{N_{evt}}$ and is
weakly dependent on $s_0$ and $\xi$ ($\sigma_\xi$ and $\hat{m}/\xi$
are roughly independent of $\xi$).  For example, from 250000 $\tau
\rightarrow 3\pi\nu_\tau$ events with $\beta_\tau \approx 0$ at the
$\tau$-charm factory, one finds $\sigma_{\hat{m}} \approx$ 1 MeV for
${\rm 2~GeV^2} \le s_0 \le m_\tau^2$.  The standard deviation
$\sigma_{\hat{m}}$
 (in
MeV) is displayed in the table below as a function of the number of
bins $N$ and of $\xi$:
\begin{quasitable}
\begin{tabular}{c|ccccc}\hline
$~$ & N=1 & N=3 & N=10 & N=36 & N=$\infty$ \\ \hline
$\xi$=0.5 & 0.5 & 0.6 & 0.8 & 0.9 & 0.9 \\
$\xi$=1.0 & 0.6 & 0.7 & 0.9 & 0.9 & 1.0 \\
$\xi$=2.0 & 0.6 & 0.7 & 0.9 & 1.0 & 1.0 \\ \hline
\end{tabular}
\end{quasitable}

It is important to realize that
$\sigma_{\hat{m}}$ increases rapidly with $\beta_\tau$, i.e., with
beam energy, because of the factor $\lambda_1$ (see Eq.\,(20) and
Fig.\,3).  Thus, the design of an experiment to measure the quark mass
must be optimized with respect to the competing effects of increasing
$\tau$ production and decreasing sensitivity as the beam energy is
raised from threshold.

In the preceding analysis, no use was made of any information coming
from measurement of the decay angle $\theta$.  In the computation of
$\sigma_{\hat{m}}$, before integration of $\theta$, $\partial \Gamma
/\partial \xi$ is proportional to $\cos \psi$.  If one makes no use of
the measurement of $\theta$, then one first integrates $\Gamma$ over
$\theta$, so $(\partial \Gamma/\partial \xi)^2$ is proportional to
$(\lambda_1)^2 = (\int \cos \psi)^2$.  As pointed out in Sec.\,3a, for
$\beta_\tau$ = 0 there is no loss in sensitivity from this procedure;
however, this is no longer true for $\beta_\tau \ne $ 0, and for
$\beta_\tau \approx 1$ (characteristic of LEP and CESR) we find that
$\sigma_{\hat{m}}$ is increased by about a factor of 4.  So, for
example, with $\rm 10^5$ $\tau \rightarrow 3\pi\nu_\tau$ events with
$\beta \approx $1 at CESR, $\sigma_{\hat{m}} \approx $ 6 MeV.

On the other hand, if one makes use of the measurement of $\theta$,
then one integrates $(\partial \Gamma/\partial \xi)^2/\Gamma$ over
$\theta$ in order to estimate the sensitivity.  One can see that this
will lead to a much less dramatic degradation of sensitivity:
ignoring the relatively weak $\theta$ dependence of the $1/\Gamma$
factor, the $\theta$ integral is proportional to $\int \cos^2 \psi =
(2\lambda_2 + 1)/3$.  A detailed study of this question is beyond the
scope of the present work.

{\flushleft{\bf 5.  OTHER COMBINATIONS OF LIGHT QUARK MASSES}}

In principle, exclusive $\tau$ decays allow the measurement of the
divergences of all four currents $\bar{d}\gamma_\mu \gamma_5 u$,
$\bar{s}\gamma_\mu \gamma_5 u$, $\bar{d}\gamma_\mu u$, and
$\bar{s}\gamma_\mu u$ which appear in Eqs.\,(2a-d).  This suggests
that the method described in detail in the previous section for the
case $m_d + m_u$ could be extended to other combinations of LQM:
$m_s + m_u$, $m_s - m_u$ and $m_d - m_u$.  We shall briefly comment on each
of these cases.

{\flushleft \bf 5a. {\boldmath $m_s + m_u$}}

Equation (2b) seems at first sight to allow a straightforward extension
of the measurement of $m_d+ m_u$ to $m_s + m_u$.  The corresponding $J$ = 0
spectral function defined as in Eq.\,(4) receives contributions from single
$K$ state and the continuum starts with $K\pi\pi$.  The corresponding
component $\rho_{K\pi\pi}$ of the spectral function could, in principle,
be measured in $\tau \rightarrow \nu_\tau + K\pi\pi$ decays.  The Cabibbo
suppression of the latter may be partially compensated by a considerably
larger ratio of signal to background ($J$ = 0 to $J$ = 1), which
is due to $m_s \gg \hat{m}$.  Unfortunately, this decay receives an anomaly
contribution $V_z$ from the $vector$ current in addition to the three
usual axial-vector form factors $A_x,A_y$ and $A_t$.  The presence of
$V_z$ essentially complicates the reconstruction of $W_{tt}$ and of
$\rho_{K\pi\pi}$ from observable quantities.  Actually, this
reconstruction turns out to be possible only using a polarized $\tau$
beam.\cite{km}  (This example emphasizes once more the {\em lucky
circumstances} which make the measurement of $\rho_{3\pi}$ and of $m_d
+ m_u$ possible.)

{\flushleft\bf 5b.  {\boldmath $m_s - m_u$}}

The difference $m_s - m_u$ controls the divergence of the {\em vector
current} $\bar{s}\gamma_\mu u$ -- cf. Eq.\,(2d).  There is no single-particle
contribution to the corresponding $J$=0 spectral function and the continuum
starts with the $K\pi$ state.  The component $\rho_{K\pi}(Q^2)$ can be measured
in exclusive $\tau$ decays
\begin{equation}
\tau^- \rightarrow K^-\pi^0 + \nu_\tau,~~~~~~~~
\tau^- \rightarrow \bar{K}^0\pi^- + \nu_\tau
\end{equation}
described by two form factors representing analytic continuations of the
well-known $K_{\ell 3}$ form factors $f_{\pm}(Q^2)$.  The $J$ = 0 and $J$ = 1
combinations of $f_{\pm}$ can be separated by measuring the polar angle
between $\vec{n}_L$ and the direction of the $K$ (or $\pi$) momentum in
the hadronic center of mass frame.  Notice that $\rho_{K\pi}(Q^2)$ -- which
is not expected to be as small as $\rho_{3\pi}(Q^2)$, since $m_s \gg \hat{m}$
-- should now be measured $directly$.  (Since there are only $two$ form
factors, it is impossible to reconstruct $\rho_{K\pi}$ from the $s-p$
interference.  The latter is, however, interesting in itself; it provides
model-independent information on $K-\pi$ phase shifts.\cite{beld})  Finally,
a lower bound for $m_s - m_u$ is provided by the sum rules\cite{ms-mu}
analogous to (8).

{\flushleft\bf 5c. {\boldmath $m_d - m_u$}}

The leading contribution to the spectral function including the divergence
(2c) of the vector current $\bar{d}\gamma_\mu u$ comes from the $\eta\pi$
state.\cite{pich}  The kinematics of the decay
\begin{equation}
\tau^- \rightarrow \pi^-\eta + \nu_\tau
\end{equation}
is completely analogous to the decays (54).  However, in this case,
the $J$=0 form factor containing the information about $m_d - m_u$
has no particular reason to be suppressed with respect to the $J$=1
form factor.\cite{pich}  The latter also vanishes as
$m_d-m_u \rightarrow 0$, and is given by isospin
breaking effects such as $\eta-\pi^0$ mixing.\cite{gl85}  The decay (55)
is obviously rare, but its observation and study would provide new
information both about $m_d-m_u$ and about the ratio
$(m_d-m_u)/(m_s-\hat{m})$.

{\flushleft\bf 6. {\boldmath CONCLUSION}}

In the present paper, we have argued that it may be possible to obtain an
experimental determination of the running quark mass $\hat{m} = (m_d + m_u)/2$
based on the measurement of angular asymmetries at the 1\% level in the decay
$\tau \rightarrow \nu_\tau + 3\pi$. The hard core of our argument is the trick,
described in Section 3b, which allows to reconstruct the dominant 3$\pi$
component ${\rho}_{3\pi}({Q^2})$ to the order $O(\hat{m}^2)$ spectral function
${\rho}(Q^2)$, associated with the divergence of the axial current, from the
measurement of the order $O(\hat{m})$ angular asymmetries.

Combining ${\rho}_{3\pi}({Q^2})$ as extracted from experiment with QCD sum
rules leads to a lower bound for the running quark mass $\hat{m}({\mu})$ as
shown by the inequality of Eq.(41). The appearance of only a lower bound arises
from the following: We have restricted our attention to the exclusive
contributions to ${\rho}(Q^2)$ coming from one and three pion intermediate
states only. Other contributions, such as $K\bar{K}\pi$ or 5$\pi$ intermediate
states, could also be studied experimentally, but they are expected to be small
due to the lack of phase space. Also, in reconstructing ${\rho}_{3\pi}({Q^2})$
we worked with a finite number of bins in the fixed $Q^2$ Dalitz plot, just as
one does in practice. Again, the question of convergence with increasing number
of bins can be addressed experimentally.

In order both to study the different steps of the above method and to get an
order of magnitude estimate of the quantities we are looking for, we have
generated a set of data from a theoretical model described in Section 4. We
wish to stress that the model on which these estimates are based should not be
given any further significance beyond its illustrative purpose. We have found
that a sample of about 250,000 background free $\tau \rightarrow \nu_\tau +
3\pi$ events is needed in order to reduce the statistical error on the lower
bound for $\hat{m}$ down to 1 Mev. This error becomes then comparable to the
theoretical uncertainties associated with the sum rules analysis.

As it stands, our analysis is mainly suited for the low energy machines, such
as a Tau/Charm Factory: In this case it is possible to integrate over the
$\tau$ decay angle $\theta$ without loss of sensitivity for the measured
angular asymmetries. If, on the other hand, one would like to use a high energy
source, such as LEP, CESR, B-factories, the method described above would have
to be adjusted and include the measurement of the $\tau$ decay angle $\theta$,
in order to compensate for the loss of sensitivity.

In the present study, we have focused on the mass combination $m_u + m_d$,
which is rather special, due to the particular role of G-parity. It also
remains the most interesting combination from the point of view of its
theoretical and phenomenological implications. In contrast, $m_s + m_u$ cannot
be given a lower bound via the same analysis, because the vector current
contributes an additional form factor to the corresponding
$\tau \rightarrow \nu_\tau + K\pi\pi$ decays. On the other hand, the quark mass
differences $m_s - m_u$ and $m_d - m_u$ can be investigated in the
quasi-two-body decays $\tau \rightarrow \nu_\tau + K\pi$ and in the rare decays
$\tau \rightarrow \nu_\tau + \eta\pi$, respectively.

We conclude that a high statistics sample of tagged, exclusive $\tau$decays
($\tau \rightarrow \nu_\tau + \pi^- \pi^- \pi^+, \pi^0 \pi^0 \pi^-, K \bar{K}
\pi^-,...$) would qualify as an authentic {\it light quark mass spectrometer}.

\acknowledgments

We are grateful to A. Roug\'e and F. Le Diberder for many valuable suggestions
and for a maximum amount of patience an experimentalist can have with a
theoretician. One of us (N.H.F.) thanks R. Decker, K.K. Gan, I.Shipsey and Z.
Was for informative correspondance. Finally, J.S. thanks H. Leutwyler for a
challenging scepticism which has triggered this study.

\pagebreak
\begin{figure}
\vspace{10cm}
\caption{Exclusive hadronic $\tau$ decays.}
\end{figure}
\begin{figure}
\vspace{10cm}
\caption{The oriented center of mass hadronic plane.}
\end{figure}

\pagebreak

\topmargin 10cm

\begin{figure}
\caption{(a) The function $\lambda_1(Q^2,\beta)$ as a function of
$Q^2/m_\tau^2$ for selected values of $\beta$.}
\end{figure}
\addtocounter{figure}{-1}
\begin{figure}
\vspace{10cm}
\caption{(b) The function $\lambda_2(Q^2,\beta)$ as a function of
$Q^2$ for $\beta$ = 1.}
\end{figure}

\pagebreak
\topmargin 10 cm
\begin{figure}
\caption{(a) The asymmetry coefficients $A,B$ as functions of $Q^2$ (in GeV).}
\end{figure}
\addtocounter{figure}{-1}
\begin{figure}
\vspace{10cm}
\caption{(b) The asymmetry coefficient $C_{LR}$
as a function of $Q^2$ (in GeV), for $\xi$ = 0.5, 1 and 2.}
\end{figure}

\pagebreak
\topmargin 10 cm
\addtocounter{figure}{-1}
\begin{figure}
\vspace{10cm}
\caption{(c) The asymmetry coefficient $C_{UD}$
as a function of $Q^2$ (in GeV), for $\xi$ = 0.5, 1 and 2.}
\end{figure}
\begin{figure}
\vspace{10cm}
\caption{(a) $W_{xx},W_{xy},W_{yy}$ integrated over the entire Dalitz plot,
for $\xi=1$, as a function of $Q^2$ (in GeV).}
\end{figure}

\pagebreak
\topmargin 10 cm
\addtocounter{figure}{-1}
\begin{figure}
\caption{(b) $W_{xt},W_{yt},W_{tt}$ integrated over the entire Dalitz plot,
for $\xi=1$, as a function of $Q^2$ (in GeV).}
\end{figure}

\begin{figure}
\vspace{10cm}
\caption{How $\sum_{i=1}^{N}\overline{W}_{tt}(Q^2,\Delta_i)$ approaches
$W_{tt}(Q^2)$ as the number of bins $N$ grows.}
\end{figure}

\pagebreak
\topmargin 10 cm
\begin{figure}
\caption{(a) Lower bounds $\hat{m}_N(\mu,s_0)$ (in MeV) as a function
of $s_0$ (in $\rm GeV^2$) and for selected values of $N$, for fixed
$\mu = 1$ GeV and $\xi$ = 0.5.}
\end{figure}
\addtocounter{figure}{-1}
\begin{figure}
\vspace{10cm}
\caption{(b) Lower bounds $\hat{m}_N(\mu,s_0)$ (in MeV) as a function
of $s_0$ (in $\rm GeV^2$) and for selected values of $N$, for fixed
$\mu = 1$ GeV and $\xi$ = 1.0.}
\end{figure}

\pagebreak
\topmargin 10 cm
\addtocounter{figure}{-1}
\begin{figure}
\caption{(c) Lower bounds $\hat{m}_N(\mu,s_0)$ (in MeV) as a function
of $s_0$ (in $\rm GeV^2$) and for selected values of $N$, for fixed
$\mu = 1$ GeV and $\xi$ = 2.0.}
\end{figure}

\end{document}